\documentclass[12pt]{article} 
\usepackage{amsmath} 
\usepackage{amssymb} 
\usepackage{graphicx} 
\usepackage[dvips]{color} 
\usepackage{colordvi}

\begin{document} 
\thispagestyle{empty} 
\addtocounter{page}{-1} 
\vskip-0.35cm 
\begin{flushright} 
TIFR/TH/09-12\\ 
%{\tt hep-th/******} 
\end{flushright} 
\vspace*{0.2cm} 
\centerline{\Large \bf Asymptotically free four-fermi theory in 4 dimensions} 
\centerline{\Large \bf at the z=3 Lifshitz-like fixed point}
\vspace*{0.2cm} 
\vspace*{1.0cm}  
\centerline{\bf Avinash Dhar$^*$, Gautam Mandal$^*$ and 
Spenta R.~Wadia$^{*\dagger}$} 
\vspace*{0.7cm} 
\centerline{\it $^*$Department of Theoretical Physics} 
\vspace*{0.2cm} 
\centerline{\it and}
\vspace*{0.2cm} 
\centerline{\it $^\dagger$International Centre for Theoretical Sciences,}
\vspace{0.2cm}
\centerline{\it Tata Institute of Fundamental Research,}  
\vspace*{0.2cm} 
\centerline{\it Mumbai 400 005, \rm INDIA} 

\vspace*{0.5cm} 
\centerline{\tt email: adhar, mandal, wadia@theory.tifr.res.in} 
 
\vspace*{0.8cm} 
\centerline{\bf Abstract} 
\vspace*{0.3cm} 
\vspace*{0.5cm} 

We show that a Nambu$-$Jona-Lasinio type four-fermion coupling at the
$z=3$ Lifshitz-like fixed point in $3+1$ dimensions is asymptotically
free and generates a mass scale dynamically. This result is
nonperturbative in the limit of a large number of fermion species. The
theory is ultra-violet complete and at low energies exhibits Lorentz
invariance as an emergent spacetime symmetry. Many of our results
generalize to $z=d$ in odd $d$ spatial dimensions; $z=d=1$ corresponds
to the Gross-Neveu model. The above mechanism of mass generation has
potential applications to the fermion mass problem and to dynamical
electroweak symmetry breaking.  We present a scenario in which a
composite Higgs field arises from a condensate of these fermions,
which then couples to quarks and leptons of the standard model. Such a
scenario could eliminate the need for the Higgs potential and the
associated hierarchy problem. We also show that the axial anomaly
formula at $z=3$ coincides with the usual one in the relativistic
domain.

\baselineskip=15pt 
 
\def\gap#1{\vspace{#1 ex}} 
\def\be{\begin{equation}} 
\def\ee{\end{equation}} 
\def\bal{\begin{array}{l}} 
\def\ba#1{\begin{array}{#1}}  %% e.g. \ba{cc} 
\def\ea{\end{array}} 
\def\bea{\begin{eqnarray}} 
\def\eea{\end{eqnarray}} 
\def\beas{\begin{eqnarray*}} 
\def\eeas{\end{eqnarray*}} 
\def\del{\partial} 
\def\eq#1{(\ref{#1})} 
\def\fig#1{Fig \ref{#1}}  
\def\re#1{{\bf #1}} 
\def\bull{$\bullet$} 
\def\mm{&&\kern-18pt}  %\mm = my marker :)-
\def\nn{\\\nonumber} 
\def\ub{\underbar} 
\def\nl{\hfill\break} 
\def\ni{\noindent} 
\def\bibi{\bibitem} 
\def\ket{\rangle} 
\def\bra{\langle} 
\def\vev#1{\langle #1 \rangle}  
\def\lsim{\stackrel{<}{\sim}} 
\def\gsim{\stackrel{>}{\sim}} 
\def\mygsim{\lower .7ex\hbox{$\gsim$}}
\def\mattwo#1#2#3#4{\left( 
\begin{array}{cc}#1&#2\\#3&#4\end{array}\right)}  
\def\tgen#1{T^{#1}} 
\def\half{\frac12} 
\def\floor#1{{\lfloor #1 \rfloor}} 
\def\ceil#1{{\lceil #1 \rceil}} 
\def\slash#1{{#1}\kern-8pt/\kern2pt}

 \def\mysec#1{\gap1\ni{\bf #1}\gap1} 
 
\def\bit{\begin{item}} 
\def\eit{\end{item}} 
\def\benu{\begin{enumerate}} 
\def\eenu{\end{enumerate}} 
%%%%%%%%%%%%%%%%%%%%%%%%%%%%%%%%%%%%%%%%%%%%%%%%%%%%%%%%%%%%%%%%%%%
\def\a{\alpha} 
\def\as{\asymp} 
\def\ap{\approx} 
\def\b{\beta} 
\def\bp{\bar{\partial}} 
\def\cA{{\cal{A}}} 
\def\cD{{\cal{D}}} 
\def\cL{{\cal{L}}} 
\def\cP{{\cal{P}}} 
\def\cR{{\cal{R}}} 
\def\da{\dagger} 
\def\de{\delta} 
\def\e{\eta} 
\def\ep{\epsilon} 
\def\eqv{\equiv} 
\def\f{\frac} 
\def\g{\gamma} 
\def\G{\Gamma} 
\def\h{\hat} 
\def\hs{\hspace} 
\def\i{\iota} 
\def\k{\kappa} 
\def\lf{\left} 
\def\l{\lambda} 
\def\la{\leftarrow} 
\def\La{\Leftarrow} 
\def\Lla{\Longleftarrow} 
\def\Lra{\Longrightarrow} 
\def\L{\Lambda} 
\def\m{\mu} 
\def\na{\nabla} 
\def\nn{\nonumber\\} 
\def\om{\omega} 
\def\O{\Omega} 
\def\p{\phi}
\def\tp{{\tilde \pi}}
\def\P{\Phi} 
\def\pa{\partial} 
\def\pr{\prime} 
\def\r{\rho} 
\def\ra{\rightarrow} 
\def\Ra{\Rightarrow} 
\def\ri{\right} 
\def\s{\sigma} 
\def\ts{{\tilde \sigma}}
\def\sq{\sqrt} 
\def\S{\Sigma} 
\def\si{\simeq} 
\def\st{\star} 
\def\t{\theta} 
\def\ta{\tau} 
\def\ti{\tilde} 
\def\tm{\times} 
\def\tr{\textrm} 
\def\T{\Theta} 
\def\up{\upsilon} 
\def\Up{\Upsilon} 
\def\v{\varepsilon} 
\def\vh{\varpi} 
\def\vk{\vec{k}} 
\def\vp{\varphi} 
\def\vr{\varrho} 
\def\vs{\varsigma} 
\def\vt{\vartheta} 
\def\w{\wedge} 
\def\z{\zeta} 
\def\psd{{\psi^\dagger}}
\def\Psd{{\Psi^\dagger}}
%%%%%%%%%%%%%%%%%%%%%%%%%%%%%%%%%%%%%%%%%%%%%%%%%%%%%%%%%%%%%%%%%%% 
\newpage

\tableofcontents 

\section{Introduction and Summary}

A fundamental problem of particle physics is the question of mass
generation of elementary particles in $3+1$ dimensions. Early attempts
in this direction were made in \cite{Nambu:1960tm, Nambu:1961tp} based
on an analogy with the theory of superconductivity. In the Standard
Model this problem is addressed by introducing the Higgs mechanism and
Yukawa couplings. The technicolour models were invented to generate
particle masses dynamically. However these have not been
phenomenologically viable for a number of reasons (see, e.g.,
\cite{Lane:2002wv,Chivukula:2000mb}).

In this paper we make an observation which has a bearing on this
question. We show that if we are willing to give up Lorentz invariance
in the ultra-violet then it is possible to have a renormalizable model
involving a Nambu$-$Jona-Lasinio type \cite{Nambu:1961tp} 4-fermi
interaction in $3+1$ dimensions. In fact, it turns out that this model
is asymptotically free and has dynamical mass generation\footnote{It
is important to note that in 4D theories involving relativistic
fermions, it is impossible to achieve asymptotic freedom without
dynamical gauge fields \cite{Coleman:1973sx}. We are able to
circumvent this theorem here by working with a Lorentz non-invariant
theory.}. Moreover, the relativistic Dirac theory emerges at low
energies. Our calculations are non-perturbative in the limit of a
large number of fermion species.

The idea that a relativistic theory at low energies may have a Lorentz
non-invariant uv-completion has been suggested recently in
\cite{Horava:2008jf,Horava:2009uw}, where the theory at high energy is
characterized by an anisotropic scaling exponent $z$ which describes
different scaling of space and time: $x \to b\,x, t \to b^zt$.
Quantum critical systems with anisotropic scaling are known in
condensed matter physics (see, e.g., \cite{Hornreich:1975zz,
Ardonne:2003wa, dimer}). Recently these theories have been discussed
in the context of AdS/(non)-CFT duality; see, e.g. \cite{Son:2008ye,
Balasubramanian:2008dm, Kachru:2008yh,Azeyanagi:2009pr}.  The idea of
relinquishing relativistic invariance at high energies has also
appeared in cosmology, e.g. as an explanation of ultra-high energy
cosmic rays above the GZK cut-off \cite{Moffat:2002nu}.  In a somewhat
different approach to the subject, Lorentz symmetry breaking has also
been used as a regulator for quantum field theories; see
\cite{Visser:2009fg} for a recent reference; see also
\cite{Pavlopoulos:1969xc}.  Currently there is a lot of interest in
the application of such ideas to gravity; however, in this paper we
will only focus on non-gravitational theories.

The plan of this paper is as follows. In Section \ref{sec:asym} we
present the 4-fermi model with $z=3$ scaling in 3 spatial
dimensions. The fermions carry a species index $i$ which takes $N$
different values. We use the large $N$ limit and compute the
nonperturbative ground state characterized by a fermion condensate. A
mass scale is dynamically generated and the 4-fermi coupling, in this
vacuum, exhibits asymptotic freedom. This result can be extended to
$z=d$ in any odd $d$ spatial dimensions. Calculations in this section
are similar to those of the Gross-Neveu model \cite{Gross:1974jv},
which can be regarded as the $z=d=1$ case. In Section \ref{sec:fluct}
we consider $1/N$ fluctuations around the condensate and show that the
phase of the condensate appears as a Nambu-Goldstone boson. When the
broken symmetry is gauged, the Nambu-Goldstone boson is `eaten up' by
the dynamical gauge field, as in the usual Higgs mechanism. In Section
\ref{sec:lorentz} we add a relevant coupling to the $z=3$ model and
discuss how a Lorentz-invariant theory emerges at low energies.  In
Section \ref{sec:anomalies} we discuss the structure of axial anomalies
in this theory and compute the anomaly coefficient.  In Section
\ref{sec:pheno} we briefly discuss application of this mechanism to
dynamical electroweak symmetry breaking. We conclude in Section
\ref{sec:discuss} with some discussions. Appendix \ref{step-gap}
provides some details of the gap equation while appendix
\ref{step-2pt} computes one-loop propagators for the bosonic
fluctuations.

Note added in v2: After the first version of this preprint was
submitted to the archive, we became aware of the work
\cite{Anselmi:2009vz} which has overlaps with the contents of this
paper.

\section{\label{sec:asym}Asymptotic freedom}

Our model consists of 2N species of fermions $\psi_{ai}(t,\vec x)$,
$a=1,2; i=1,..., N$ which carry representations of $SU(N)$ and a
flavour group $U(1)_1 \times U(1)_2$, as follows:
\bea 
&&\psi_{ai} \to U_{ij}\psi_{aj}
\nn 
&& \psi_{ai} \to e^{i\alpha_a} \psi_{ai}, \; a=1,2 
\eea 
Each of these fermions is an $SU(2)_s$ spinor, where $SU(2)_s$ is
the double cover of the spatial rotation group $SO(3)$. 

An action which is consistent with the above symmetries is:
\bea
S &&= \int d^3\vec  x\ dt \left[
 \psd_{\!1i} 
\left(i \del_t + i \vec \del. \vec \s\ \del^2\right) \psi_{1i}
+  \psd_{2i}\left(i \del_t - i \vec \del. \vec \s\ \del^2\right)\psi_{2i}
 \right.
\nn
&& 
\left. \kern+100pt + g^2 \ \psd_{1i}\psi_{2i} \psd_{2j} \psi_{1j} \right],
\label{four-fermi}
\eea
where $\{\vec \sigma\}$ are the Pauli matrices. We will study 
the dynamics of this action in the large $N$ limit in which $\l= g^2
N$, the 'tHooft coupling, is held fixed. Note the sign flip of
the spatial derivative term between the two flavours $a=1$ and $a=2$;
this ensures that the Lagrangian is invariant under a parity operation
under which $\psi_{1i}(t, \vec x) \to \psi_{2i}(t,-\vec x) $.

Note that if we assign scaling dimensions according to $z=3$, i.e. $
[L]= -1, [T]=-3$ , then $[\psi]= 3/2$. In this case, all the three
terms appearing in the above action are of dimension 6 and hence
marginal.

It is important that the four-fermion interaction term is marginal at
$z=3$.  Recall that in the usual context of a $3+1$ dimensional
Lorentz invariant theory, any interaction involving four fermions
represents an irrelevant operator and so must be understood as a low
energy effective interaction. By contrast, here the marginality of the
interaction leads one to hope that the theory \eq{four-fermi} is
perhaps uv-complete. We will show below that this is indeed the case
since the four-fermi coupling turns out to be asymptotically free.

A more general $z=3$ action which considers all relevant and
marginal couplings, and is consistent with
the symmetries of \eq{four-fermi},  can also be written down: 
\bea
\!\!\!S =\!\! \int\!\!\mm  d^3\vec  x\  dt \left[ 
 \psd_{\!1i} 
\left(i \del_t - i \vec \del. \vec \s\ \left((-i\del)^2+ g_1\right)
+ g_2 \del^2 \right) \psi_{1i} 
\right.
\nn
\mm
+  \psd_{\!2i} 
\left(i \del_t + i \vec \del. \vec \s\ \left((-i\del)^2+ g_1\right)
+ g_2 \del^2 \right) \psi_{2i}
+  g_3 \left( \psd_{\!1i} \psi_{1i} +  
\psd_{\!2i} \psi_{2i} \right)  
\nn
\mm
\left.
 + g_4^2 \left(\left(\psd_{\!\!1i}\psi_{1i}\right)^2 + 
\left(\psd_{\!\!21} \psi_{2i}\right)^2\right)
+ g_5^2 \left( \psd_{\!\!1i}\psi_{1i} \psd_{2j}\psi_{2j} \right)
+ g_6^2 \left( \psd_{\!\!1i}\psi_{2i} \psd_{2j}\psi_{1j} \right)
 \right],
\nn
\label{relevant}
\eea
The action \eq{four-fermi} corresponds to putting all the
couplings $g_1,..., g_5=0$ and setting $g_6= g$. We will
treat some of these other couplings in Section \ref{sec:marginal}
and Section \ref{sec:lorentz}.

One can eliminate the four-fermi interaction in \eq{four-fermi} by
using a standard Gaussian trick:
\[
\exp\left[ i \left( g^2 \!\!
\int\! \psd_{1i}\psi_{2i} \psd_{2j} \psi_{1j}
\right)\right] 
=
\int {\cal D}\phi \ \exp\left[i \int\!  \p^* \psd_{1i}\psi_{2i}
+ \ \p \psd_{2i}\psi_{1i} - \frac1{g^2} \p^*\p \right]
\]
This gives us the following action, which is entirely
equivalent to \eq{four-fermi}:
\bea
S \mm = \int d^3\vec  x\ dt \left[
 \psd_{1i} \left(i \del_t + i \vec \del. \vec \s \del^2\right) \psi_{1i}
+  \psd_{2i}\left(i \del_t - i \vec \del. \vec \s \del^2\right)\psi_{2i}
\right.
\nn
&& 
+ \left.  \p^* \psd_{1i}\psi_{2i}
+ \p \psd_{2i}\psi_{1i} - \frac1{g^2} \p^*\p \right]
\label{auxil}
\eea
The scalar field $\p$ is an $SU(N)$-singlet and is charged 
under the axial $U(1)$ parametrized by $\exp[i(\alpha_1 - \alpha_2)]$.

\subsection{\label{sec:gap}The gap equation}

Since the action \eq{auxil} is quadratic in fermions, one can
integrate them out, leading to the following effective action for the
boson:
\be
S_{\rm eff}[\phi] 
= -i N {\rm Tr} \ln \tilde D  - \frac1{g^2} \int \p^*\p
\label{s-eff}
\ee
where $\tilde D$ is defined in \eq{tilde-d}.  Here ${\rm Tr}$
represents a trace over space-time as well as the flavour and spinor
indices.

In the large $N$ limit, the classical equation of motion 
$\delta S_{\rm eff}/\delta \phi=0$ is exact, leading to
(see Section \ref{step-gap} for details)
\be
i\int \frac{d^4k}{(2\pi)^4}\frac1{k_0^2 - |\vec k|^6 - \p^*\p + i\ep}
=   \frac{1}{2\l},\  \l= g^2 N 
\label{gap}
\ee
This gap equation determines only the absolute value of
$\p$. The phase of $\p$ will be identified with a Nambu-Goldstone
mode $\pi$ in the next section where we consider fluctuations.

The left-hand-side of the gap equation is logarithmically divergent by
$z=3$ power counting (both numerator and denominator have dimension
6). Rotating the contour from $k_0 \in (-\infty, \infty)$ to
$k_0\in (-i\infty, i\infty) $ (this is an anticlockwise rotation in
the complex $k_0$ plane by $\pi/2$ and can be done without touching
the poles of the Feynman propagator), we get
\be
\int \frac{dk_0 d^3k}{(2\pi)^4}\frac1{k_0^2 + k^6 + \p^*\p}
=   \frac{1}{2\l} 
\label{gap-eucl}
\ee
It is easy to do the angular integration. Then, using the 
variable $w = k^3$ and extending the range of $w$-integral to the
entire real line (possible because the integrand has $w
\leftrightarrow -w$ symmetry), we get
\be
\frac{2\pi/3}{(2\pi)^4}\int_{-\infty}^\infty dk_0 
\int_{-\infty}^\infty dw\ \frac1{k_0^2 + w^2 + \p^*\p}
=   \frac{1}{2\l} 
\label{k0-w}
\ee
The above integral has an $SO(2)$ rotational symmetry between
$k_0$ and $w$. In particular, if we parametrize
\be
(k_0, w) = K (\cos\theta, \sin\theta), \theta \in [0, \pi]
\label{so2}
\ee
then the angle $\theta$ can be integrated out. Using the
$SO(2)$-invariant cutoff $K \le \L^3$ and discarding
a finite piece \footnote{The actual result for the LHS using this
cut-off is $\ln(1+ \L^6/|\p|^2)$. The finite pieces depend
on the cut-off scheme, e.g.  if, in \eq{k0-w}, we integrate $k_0$
first from $-\infty$ to $\infty$ and then $w$ from $0$ to $\L^3$,
the LHS of \eq{gap-final} would be $\ln(\sqrt{1+ \L^6/|\p|^2}+
\L^6/|\p|^2)$.}, we get
\be
\ln \left(\frac{\L}{m} \right)= \frac{2\pi^2}{\l}
\label{gap-final}
\ee 
where $\L$ has momentum dimension one and we have introduced the 
parameter $m$ of momentum dimension one by defining $m^6
\equiv \p^* \p$. In the 
large $N$ limit, fluctuations of $\p$ are suppressed and this 
solution of the gap equation becomes exact.  

We see from \eq{auxil} that, around this symmetry broken vacuum, the
term involving the parameter $m$ is like a mass term for the fermions.
When we perturb this model by adding a relevant term that takes it at
low energies to the relativistic fixed point at $z=1$, this term goes
over to the familiar mass term for relativistic fermions, with a mass
proportional to $m$. This is discussed further in Sec.4.

Eqn. \eq{gap-final} determines $m$ in terms of $\l$ and
the cut-off $\L$. We will demand that $\l$ must be assigned an
appropriate $\L$-dependence such that the fermion mass $m^3 = |\vev \p|$
is kept invariant. From \eq{gap-final} this gives us
\be
\l(\L)  = \frac{2\pi^2}{\ln(\L/m)}
\label{asym}
\ee
We see that $\l$ is an asymptotically free coupling. The theory
generates a mass scale analogous to $\L_{\rm QCD}$, given by
\[
m= \L  \exp[-\frac{2\pi^2}{\l}]
\]
The $\beta$-function is easy to compute and it is negative:
\[
\beta(\l)= \L\frac{d\l}{d\L}=  - \frac{\l^2}{2\pi^2}
\]
The calculation presented above is similar to that for the Gross-Neveu
model \cite{Gross:1974jv}. Indeed, we will show in the next subsection
that the results presented above generalize to all odd $d$ spatial
dimensions at $z=d$. The Gross-Neveu model, from this viewpoint, is
simply the $d=1,\, z=1$ example. Unlike in the higher dimensional
examples, however, the fermion condensate in the Gross-Neveu model
breaks only a discrete $Z_2$ symmetry and there is no Nambu-Goldstone
mode.

We should point out that the condensate is generated here for
arbitrarily weak coupling $g$. This is in contrast with what happens
in the usual relativistically invariant NJL model at the $z=1$ fixed
point \cite{Nambu:1961tp, Dhar:1983fr, Dhar:1985gh, Dhar:2009gf},
where the symmetry breaking phase occurs only beyond a certain
critical value $g_c$ of the coupling.

It is useful to express the above result in terms of an effective
potential for the homogeneous mode of $\phi$. This is essentially
the negative of \eq{s-eff}, evaluated for constant $\phi$, and is
given by 
\be
V_{\rm eff}(\phi)=  \frac{|\phi|^2}{g^2}\left(1- \frac{\l}{12\pi^2} 
(\ln\frac{\Lambda^6}{|\phi|^2} +1) \right)
\label{eff-pot}
\ee
A plot of $V$ as a function of $|\phi|$ looks like

\gap1

\centerline{\includegraphics[scale=.6]{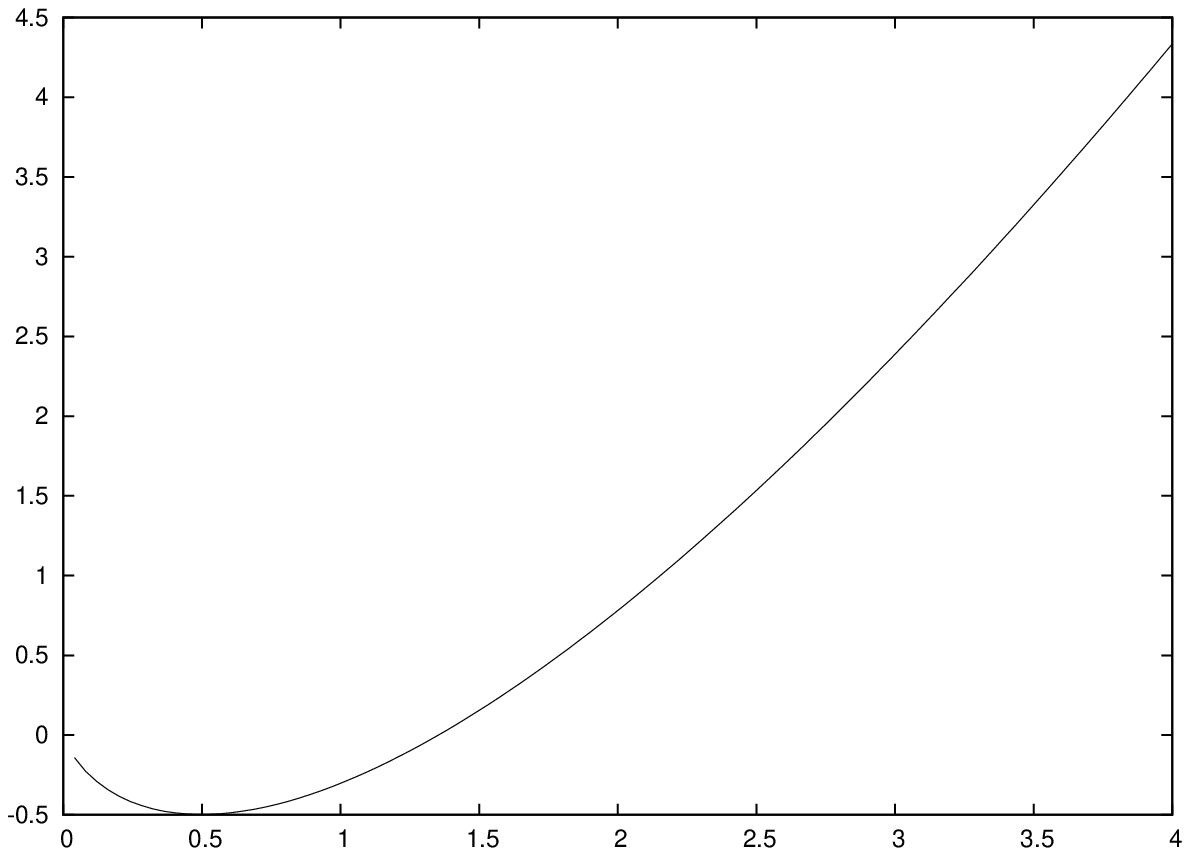}}
\gap1

\ni which shows a minimum at 
\be
|\phi| =m^3= \Lambda^3 \exp[-6\pi^2/\l], 
\label{minimum}
\ee
as found above. We emphasize that the treatment of the
effective potential and the RG flow presented above is exact
in the strict $N=\infty$ limit.   

\subsection{\label{other-d}Other dimensions and $z=d$}

In this subsection we show that the above conclusion generalizes to
$z=d$ in $d= 2n+1$ spatial dimensions. We will again consider fermions
$\psi_{ai}$ which transform in the fundamental representation of
$SU(N)$ and a flavour group $U(1)_1 \times U(1)_2$; each fermion
transforms as a spinor of (an appropriate covering group of) the
spatial rotation group $SO(2n+1)$. The action \eq{auxil} now reads:
\bea 
S && = \int d^{2n+1} \vec x\ dt \left[ \psd_{1i} \left(i \del_t
+ i \vec \del. \vec \G \del^{2n}\right) \psi_{1i} + \psd_{2i}\left(i
\del_t - i \vec \del. \vec \G \del^{2n}\right)\psi_{2i} \right.  
\nn
&& \kern100pt \left. + g^2 \ \psd_{1i}\psi_{2i} \psd_{2j} \psi_{1j} \right]
\label{four-fermi-d}
\eea
Here $\G^i, i=1,2,..,2n$ are the gamma matrices in $2n$ Euclidean
dimensions. For $z=d$, the dimension of the fermion is $[\psi]=
d/2$. Hence the 4-fermi coupling is marginal for any $d$. 

The gap equation now reads
\be 
2^{n+1} \int \frac{dk_0\ d^{2n+1}k}{(2\pi)^{2n+2}}
\frac1{k_0^2 - k^{2n+2} -
\p^*\p + i\ep} = \frac{2}{\l},
\label{gap-d}
\ee
from which we get 
\[
\l(\L)= \frac{A}{\ln(\L/m)},\ A= 2\pi^{n+1} (2n-1)!!
\]
showing asymptotic freedom of the coupling. Here $(2n-1)!!$
$=(2n-1)(2n-3)...1$ for $n\ge 1$ and $=1$ for $n=0$. The beta-function
is given by
\[
\beta(\l)= -\frac{1}{A} \l^2
\]
Note that the $\beta$-function vanishes exponentially as $d
\rightarrow \infty$.

\subsection{\label{sec:marginal} The space of marginal couplings}

In the bulk of this section, we have dealt with the action
\eq{four-fermi}, or equivalently with \eq{auxil}. It is not difficult
to generalize our results to include the marginal couplings $g_4$ and
$g_5$ in \eq{relevant}. We postpone the details to a forthcoming
publication \cite{progress} and simply quote the results here. As
before, we can use a Gaussian trick to replace the new quartic
interactions by introducing additional auxiliary scalar fields.  A
general effective potential involving these fields can be obtained
along the lines of \eq{eff-pot}.  It turns out that (a) the new terms
in the effective potential do not depend on the cut-off, and (b) at
the mimimum of the potential $\vev{\phi}$ is still given by
\eq{minimum} whereas the other condensates vanish. Furthermore, only
the coupling $g_6=g$, considered before, has a non-trivial
beta-function and the theory remains asymptotically free. If we do
include the relevant coupling $g_3$, then the other condensates
acquire non-vanishing vev's; however, even then it is only $g_6$ which
has a non-trivial RG flow and the theory is asymptotically free. We
will discuss the consequence of including the relevant coupling $g_1$
in Section \ref{sec:lorentz}.

\section{\label{sec:fluct}Quantum fluctuations}

In the previous section, we considered the classical solution of
$S_{\rm eff}(\p)$ (Eqn. \eq{s-eff}), which is exact in the large $N$
limit. In this section we will go beyond this approximation and
consider fluctuations of the scalar field $\p$. It is convenient to
parametrize the fluctuations in terms of a radial field (sigma) and a
phase (pion):
\be
\p(x) = \r(x) e^{i g\pi(x)}, \; \r(x) = m^3+ \frac{g}{\sqrt{2}} \s(x)
\label{def-pi}
\ee
It is convenient to use the notation of Dirac
matrices and rewrite the action
\eq{auxil}in the form given by \eq{auxil-gamma} and \eq{app-dirac}. 
Substituting \eq{def-pi} in these equations, we get the following 
action for fluctuations:
\bea
S&&\kern-20pt = \int d^4x \left[
\bar\Psi_i \left( i \g^0\del_t + i(\vec \g.\vec \del)(i \vec \del)^2 \right) 
\Psi_i + \bar\Psi_i \left((m^3 + \frac{g}{\sqrt{2}} \s(x)) 
e^{i g \pi(x)} P_L\right. \right.
\nn
&& \left. \left. 
+ (m^3 + \frac{g}{\sqrt{2}} \s(x)) e^{-ig \pi(x)} P_R \right) \Psi_i 
 - \frac1{g^2} (m^3 + \frac{g}{\sqrt{2}} \s(x))^2
\right]
\label{fluct-action-1}
\eea
where $P_{L,R} = \frac12 (1 \pm \g^5)$. The action has the following
global $U(1)$ symmetry
\be
\Psi_i \to e^{i g\alpha \g^5}  \Psi_i, \; \pi \to \pi - \alpha
\label{axial}
\ee
In terms of the original $U(1)_1 \times U(1)_2$ symmetry of the
action, this is the off-diagonal (axial) $U(1)$. The fermion
condensate breaks this symmetry, with the pion $\pi(x)$ as a
Nambu-Goldstone boson.

The masslessness of the pion can be argued as follows. By making a
local phase rotation $\Psi_i \to e^{-i g \g^5 \pi(x)/2 } \Psi_i$ in
the fermion functional integral, the pion field can be eliminated from
the Yukawa coupling terms, with the replacements
\be
\del_t \to \del_t + \frac{ig}{2}\del_t \pi, \;
\del_i \to \del_i + \frac{ig}{2}\del_i \pi
\label{cov-deriv}
\ee
in the fermion kinetic terms. This shows that the effective action
\eq{s-eff} contains the pion field only through its derivatives,
which, therefore, rules out a mass term.

The above argument relies on the invariance of the fermionic measure
under an axial phase rotation and could be potentially invalidated by
the appearance of anomalies. We will discuss this issue in the next
subsection which deals with coupling to gauge fields. For the present,
we note that in the absence of gauge fields any potential anomaly
vanishes and the argument about the masslessness of the pion goes
through.

In Sec. \ref{step-2pt}, further evidence for the masslessness of the
$\pi$ field is provided by an explicit computation of the one-loop
propagator for the bosonic fluctuations.

\subsection{\label{sec:gauge}Coupling to gauge fields}

If we gauge the axial $U(1)$ by appropriately coupling the fermions to
a dynamical gauge field, then the effect of the phase rotation
$\exp[-ig \g^5\pi(x)/2]$ on the fermions will be to replace the gauge-covariant
derivatives in a manner analogous to \eq{cov-deriv}. The pion field
and the gauge field will then appear in an extended covariant
derivative of the form
\be
\tilde D_t = \del_t + i e(A_t + \frac{g}{2e} \del_t \pi), \;
\tilde D_i = \del_i + i e(A_i + \frac{g}{2e} \del_i \pi)
\label{eat}
\ee

This shows that the gauge field effectively absorbs the pion field, as
in the standard Higgs mechanism, and becomes massive.  The mechanism
of generation of mass terms for the gauge field mass terms arise in a
manner This is discussed further in Sec. \ref{sec:pheno}.

As in the previous subsection, we have assumed invariance of the
fermion functional integral under a local axial phase rotation.  This
argument is valid only if there are no anomalies of the axial
current. In Section \ref{sec:anomalies} we will calculate this anomaly
and show that it is the same as for relativistic fermions. The absence
of this anomaly therefore imposes the usual requirement on the
spectrum of fermions coupled to the axial gauge field.

\section{\label{sec:lorentz}Emergence of Lorentz invariance}

In this section we will consider the effect of adding the relevant
coupling $g_1$, defined in \eq{relevant}, to the theory
\eq{fluct-action-1}.  According to $z=3$ scaling, the momentum
dimension of $g_1$ is $2$. Denoting $g_1 \equiv M^2$, the action reads
\bea
S&&\kern-20pt = \int d^4x \left[
\bar\Psi_i \left( i \g^0\del_t + i(\vec \g.\vec \del)(M^2
+(i \vec \del)^2) \right) 
\Psi_i + \bar\Psi_i \left((m^3 + \frac{g}{\sqrt{2}} \s(x)) 
e^{i g \pi(x)} P_L\right. \right.
\nn
&& \left. \left. 
+ (m^3 + \frac{g}{\sqrt{2}} \s(x)) e^{-ig \pi(x)} P_R \right) \Psi_i 
 - \frac1{g^2} (m^3 + \frac{g}{\sqrt{2}} \s(x))^2
\right]
\label{action-relevant}
\eea
The mass shell condition of the fermion is
\be
k_0^2 - k^2 (M^2 + k^2)^2 - m^6 =0
\label{mass-shell}
\ee
For the momentum range
\be
k \ll M,
\label{lorentz-range}
\ee
we get 
\[
k_0^2 - M^4 (k^2 + {m_*}^2) =0, \quad m_*=\frac{m^3}{M^2}
\]
Let us introduce a rescaled time and energy 
\be
t'= t M^2,  k_0'=  k_0/M^2
\label{rescaling}
\ee
so that $t'$ is of mass dimension $-1$ and $k_0'$ of
mass dimension 1. The mass shell condition becomes the standard
form dictated by Lorentz invariance.
\be
(k_0')^2 = (k^2 + {m_*}^2) 
\label{lorentz-shell}
\ee 
In the momentum range \eq{lorentz-range} and in terms of the
rescaled time \eq{rescaling}, the action \eq{action-relevant} becomes
\bea 
S&&\kern-20pt = \int d^3x dt' \left[ \bar\Psi_i \left( i
\g^0\del_{t'} + i\vec \g.\vec \del \right) \Psi_i + \bar\Psi_i
\left((m_* + \frac{g}{\sqrt{2}} \s'(x)) e^{i g \pi(x)}
P_L\right. \right.  \nn 
&& \left. \left.  + (m_* + \frac{g}{\sqrt{2}}
\s'(x)) e^{-ig \pi(x)} P_R \right) \Psi_i - \frac{M^2}{g^2} (m_* +
\frac{g}{\sqrt{2}} \s'(x))^2 \right]
\label{action-relevant-lorentz}
\eea 
where we have defined the rescaled bosonic field $\s'= \s/M^2$.
In view of \eq{lorentz-range}, the above action should be understood
with an effective cut-off $M$.

The following points are worth noting:

\ni 1. In principle, the coupling $g_1$ in \eq{relevant} actually
flows under RG (see the comment below \eq{generate-g1}). Hence,
strictly speaking, we should make a distinction between the
coefficient $M$ appearing in \eq{action-relevant} and the constant $M$
appearing in \eq{mass-shell}.  The former should be regarded as a
cut-off dependent coupling $M(\L)$, whereas the latter should be
regarded as an RG-invariant mass scale $M$ entering in the mass-shell
condition \eq{mass-shell}. However, in the present case the
corrections are suppressed by $1/N$ and can be neglected in the large
$N$ limit. Therefore, in the present case, the $M$ appearing in the
two equations \eq{generate-g1} and \eq{mass-shell} are the same.

\ni 2. Although the coupling $\l$ is asymptotically free and grows
logarithmically at low $k$, the coupling constant relevant to quantum
fluctuations around the fermion condensate is $g = \sqrt{\l/N}$
remains weak at large $N$.  

\ni 3. The $\psd\psi\s$ coupling $g$, remarkably, is marginal both at
$z=3$ and at $z=1$. The reason is that in \eq{action-relevant} $\s$ 
has dimension 3, hence $\psd\psi\s$ has dimension 6 (which is marginal
at $z=3$); on the other hand in \eq{action-relevant-lorentz} the
rescaled bosonic field $\s'=\s/M^2$ has dimension 1, hence
the $\psd\psi\s'$ coupling has dimension 4 (which is marginal
at $z=1$).

The various mass scales which appeared above can be schematically
represented on the momentum line:

\gap3

\centerline{\includegraphics[scale=.65]{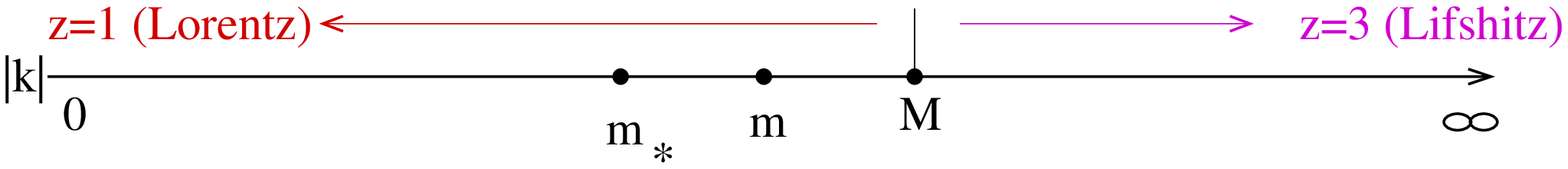}}

\ni Here $m$ is the mass scale generated by the condensate,
$m=|\phi|^{1/3}$. $M$ is the momentum scale below which Lorentz
symmetry appears. $m_*$ is the rest mass of the emergent Dirac
fermion.  The three masses are related by $m^3 = m_* M^2$.  Here we
have chosen the order $M \gg m \gg m_*$ for potential applications to
weak interaction (see Section \ref{sec:pheno}).  Theoretically,
the other order  $M \ll m \ll m_*$ is also allowed; however,
with that choice, the ``Higgs vev'' is higher than the scale of
violation of Lorentz symmetries, which is unrealistic.

\section{\label{sec:anomalies} Anomalies}

In this section we present a brief discussion of axial anomalies when
our model \eq{four-fermi} is coupled to gauge fields. We will first
consider the case where the diagonal $U(1)$ is gauged and will look
for possible anomalies in the axial $U(1)$ current.

We will compute the anomaly adapting the method of Fujikawa
\cite{Fujikawa:1979ay} (see also \cite{Dhar:1983fr,Dhar:1985gh}) to our
problem. The path integral for the gauged model is given by (using the
form of the fermion action given in \eq{action-relevant})\footnote{We
use $i$ to label the fermion as well as spatial coordintes; the
specific usage should be clear from the context.}  
\bea 
Z \mm = \int
\prod_i {\cal D}\Psi_i {\cal D} \bar\Psi_i\ e^{iS},\; S = \int d^3
x\,dt\, \bar\Psi_i i \slash{\bf D} \Psi_i 
\nn 
\slash{\bf D} \mm =
\g^\mu {\bf D}_\mu,\; {\bf D}_t = D_t, \, {\bf D}_i= D_i \left( -
(\vec D)^2 + M^2\right), \, D_\mu= \del_\mu + ie A_\mu
\label{gauged}
\eea
Here we have ignored the four-fermion term since it does not play a
role in the calculation of anomalies.  We will treat the gauge field
as an external field.  If we make a local axial rotation
\be
\delta \Psi_i = i \alpha(x) \g^5 \Psi_i,\;
\delta \bar \Psi_i = \bar \Psi_i\, i \alpha(x)  \g^5  
\label{axial-rotn}
\ee
we generate a term proportional to $\int \alpha(x) \del_\mu J^{\mu 5}$
in the action and if the path integral measure is invariant under
\eq{axial-rotn} then the axial current is conserved. However, it was
shown in \cite{Fujikawa:1979ay} that under \eq{axial-rotn}, the
measure picks up a Jacobian. Taking this into account in the present
case, we get \footnote{Note that the spatial component 
of the current, ${\vec J}^5$, is much more complicated than
its relativistic counterpart.} 
\[
\del_\mu J^{\mu 5} = 2 \langle x |
{\rm tr} \left(\g^5\exp[{i \slash{\bf D}}^2/\L^6]\right) | x \rangle  
%\label{del-j5}
\]
where the gauge-invariant expoential operator is introduced as a
regulator, as in \cite{Fujikawa:1979ay}, except that the cut-off $\L$
follows $z=3$ scaling ({\em cf.} Section \ref{sec:gap}).  Here `tr'
refers to a Dirac trace as well as a sum over the species index $i=1,
...,N$.  To evaluate the RHS, we will expand the exponential in powers
of the charge $e$, by using
\[
(\slash{\bf D})^2  =  -{\bf D}_\mu{\bf D}^\mu   +
\frac{i}2 \S^{\mu\nu}[{\bf D}_\mu,{\bf D}_\nu]   
\]
The second term is proportional to $e^2$.  In order for the Dirac
trace to survive, we must bring down two powers of this term.  
Using ${\rm tr} (\g^5 \S^{0i}\S^{jk})=4i \ep^{ijk}$
and ignoring terms which would drop out at large $\L$
(these include all terms involving $M$) we find
\beas
\del_\mu J^{\mu 5} \mm = - 
\frac{2i e^2 N}{\L^{12}} \ep^{ijk} 
\langle x | \exp[(-\del_0^2+ (\vec \del)^6)/\L^6] \
\{ E_i (\vec \del)^2 + E_p \del_p \del_i,
\nn
\mm \kern80pt
F_{jk}(\vec \del)^4 + 2 F_{jl} \del_l \del_k (\vec \del)^2
+ 2 F_{lk} \del_l \del_j (\vec \del)^2 \} | x \rangle
\eeas
The terms inside $\langle x | ... | x \rangle$ give
rise to various powers of $\L$, the highest being 
$\L^{12}$ which is exactly cancelled by the $1/\L^{12}$
outside. Terms involving derivatives of the electric or magnetic
field give rise to lower powers of $\L$ and eventually drop out.
The final result, at the end of a long calculation, is
\be
\del_\mu J^{\mu 5} = -\frac{e^2 N}{16\pi^2} 
\ep^{\mu\nu\l\s}F_{\mu\nu}F_{\l\s}
\label{anomaly-coeff}
\ee
which is identical to the usual relativistic calculation of the axial
anomaly!\footnote{We can, in fact, recover \eq{anomaly-coeff} in the
$z=1$ limit from \eq{gauged} by making the replacement $ - (\vec D)^2
+ M^2 \to M^2$.} The robustness of the anomaly coefficient with
respect to different $z$ values is likely to be related to its
topological character. The result is also gratifying from the
viewpoint of model building since we do not want to impose different
requirements on the fermion spectra at different energy scales.

The axial anomaly for chiral fermions (only $\psi_{1i}$ and no
$\psi_{2i}$) as well as chiral anomalies for chiral gauge theories
can be obtained by simple generalizations of the computation
presented above. 

\section{\label{sec:pheno}Application to low energy phenomenology}

%\enlargethispage{1000pt}
In this section we will consider a simple extension of the fermion
model \eq{auxil} which can describe electroweak symmetry breaking. The
extension consists of an additional $SU(2)$ group, under which the
$a=1$ fermions transform as a doublet and the $a=2$ fermions transform
as a singlet. Using the Dirac spinor notation employed in the previous
section, let us denote the $a=1$ fermions as $\psi_L$ (these satisfy
$\g^5=1$) and $a=2$ fermions as $\psi_R$ (these satisfy
$\g^5=-1$). The fermion fields will then be denoted as $\psi_{Li\a},
\psi_{Ri}$ where $\a=1,2$ is the new $SU(2)$ index 
\footnote{In order to generate masses for all the $\psi$-fermions, we
need to double the number of right-handed fermions as well, still
keeping them singlets under the above SU(2). In this more
general model, the four-fermi couplings
can be  arranged such that one still has only a doublet
of Higgs in the broken phase}. We then couple the
fermions to $SU(2)$ gauge fields.\footnote{We can also add a gauge
field to gauge the vector part of $U(1) \times U(1)$.}

The scalar field, $\phi_\a$, which is classically equivalent to the
fermion bilinear $g \psi_{Ri}\psi_{Li\a}$, now carries the additional
$SU(2)$ index $\alpha$ and transforms as a doublet. This will play the
role of the composite Higgs field.

In addition to the above fermions, we will have the usual quark and
lepton degrees of freedom. These do not carry the species index $i$,
but they do have quartic interaction terms with the above fermions,
similar to those in \eq{four-fermi}. These quartic interactions are
designed to respect the $SU(2)$ gauge symmetry and the global
symmetries of the action. An example is
\be
(\psd_{Li\alpha}\psi_{Ri})(q^\dagger_{R} q_{L\alpha})
\label{yukawa}
\ee
where the $q$'s denote quarks. This interaction will generate the
Yukawa couplings after the $\psi$'s have condensed, as we show
below.

We can now repeat the analysis of Sections \ref{sec:asym} and
\ref{sec:fluct} to show that $\phi_\a$ develops a vev, thereby
dynamically breaking the gauge symmetry. By parametrizing $\phi =
\exp(i\vec \pi(x).\vec \tau) \rho$, we can show, as in Section
\ref{sec:gauge}, that $\vec \pi(x)$'s combine with the $SU(2)$ gauge
fields to give them their longitudinal components. The fluctuation
$\s_\a(x)$ of the radial field $\rho_\a(x)$ becomes the massive Higgs
field, in terms of which \eq{yukawa} gives the usual coupling
of quarks to the Higgs field:
\be
q^\dagger_{R} \s_\a q_{L\alpha}
\label{higgs}
\ee
The gauge field masses arise from their gauge-invariant interactions
with the $\psi$'s. The relevant diagram is shown in the following
figure.

\gap{5}

\centerline{\includegraphics[scale=.7]{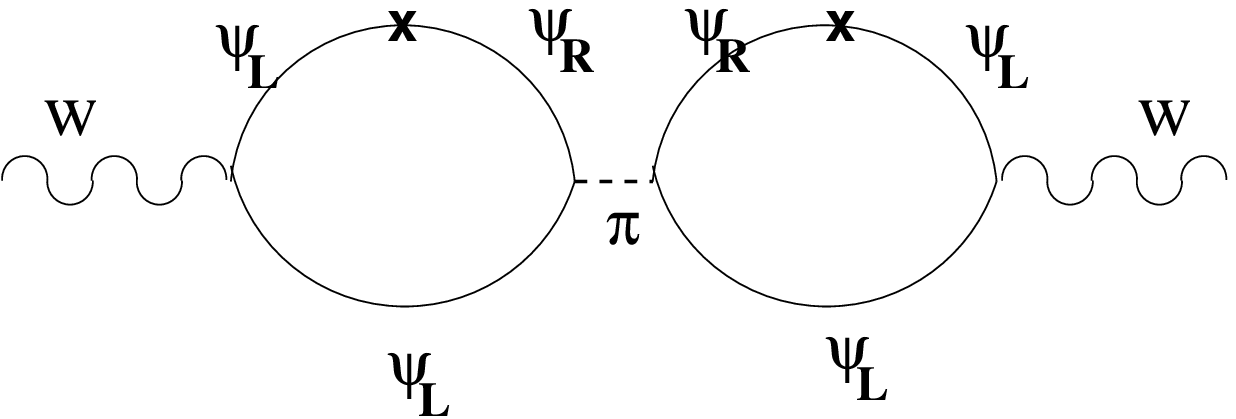}} 

\gap{3}

\ni The crosses on fermion propagators indicate insertions of the
dynamically generated mass. The main point is the exchange of the
massless Nambu-Goldstone ``pion'', which is responsible for generating
the gauge boson masses. This well-known mechanism was originally
discovered in the context of Meissner effect (see \cite{Nambu:1960tm}
and references therein).

\gap1

We conclude this section with the following observations:

%\begin{enumerate}

\gap2

%\item
\ni 1. The introduction of the fermions ($\psi_{Li\alpha}, \psi_{Ri}$)
with the marginal four-fermi coupling gives rise to a composite
Higgs field and hence eliminates the need for the Higgs
potential. Consequently, the hierarchy problem of the standard model
is avoided.

%\item
\ni 2. We used a large number $N$ of these fermions in order to make
the point that the model exists nonperturbatively and that the
fluctuations around the nonperturbative vacuum are weakly
coupled. However, this is not an in-principle requirement if we can
tackle the four-fermion coupling nonperturbatively by some other
method.

%\item
\ni 3. For realistic applications, we need to ensure that the mass
$m_*$ of the extra fermions ($\psi_{Li\alpha}, \psi_{Ri}$) is beyond
the presently observed energy scales while the Higgs mass and the
Higgs vev are not too high. We wish to come back
to a detailed analysis of this constraint in a later paper.

%\item
\ni 4. The mechanism of mass generation presented here is much simpler
than in technicolour theories \cite{Lane:2002wv, Chivukula:2000mb}.

\ni 5. The spectrum and quantum number of fermions is constrained by
the requirement of vanishing of gauge anomalies, the computation of
which is described in the Section \ref{sec:anomalies}.
 
\ni 6. Once the fermions $\psi_{Li\alpha}$ are coupled to gauge fields,
the coupling $g_1=M^2$ will get renormalized, even to leading order
in large $N$. In this case, a scale dependent $g_1$ will enter in
the fermion  mass-shell condition, and, more generally, in the
effective ``speed of light'' parameter for each particle.  In this
case, one has to make sure that the variation of the speed of light
is compatible with existing bounds from experiments.  

%\end{enumerate}

\section{\label{sec:discuss}Discussion}

In this paper, we have shown that at the $z=3$ fixed point, an
NJL-like 4-fermi coupling in $3+1$ dimensions is asymptotically free,
thus providing an uv completion of the low-energy 4-fermion coupling
at the $z=1$ fixed point.  The price to pay is Lorentz non-invariance
in the ultraviolet. Our work provides a novel composite Higgs
mechanism for dynamical gauge symmetry breaking and generation of
fermion masses. 

The asymmetry in the ultraviolet cut-off corresponding to space and
time directions may be a fundamental feature of our world. If true,
this feature could have important consequences for low energy particle
physics and model building. It would be interesting to
see if the $z=3$ model described here for the electroweak sector can
be extended to include strong and gravitational interactions,
especially in the framework of string theory. For this reason, it is
important to explore the formulation of string theory itself which
incorporates Lorentz violation in the ultraviolet. For example, we
observe that in the exact formulation of 2-dimensional string theory
in terms of matrix quantum mechanics, one naturally arrives at a $z=2$
theory of non-relativistic fermions \cite{Sengupta:1990bt,
Gross:1990st, Polchinski:1991uq, Jevicki:1979mb, Das:1990kaa}. The
theory becomes relativistic ($z=1$) only for low energy fluctuations
around the fermi surface.

\subsection*{Acknowledgement} 

We would like to thank Kedar Damle, Cesar Gomez and especially Sreerup
Raychaudhuri for many valuable discussions. We also wish to thank
Manav Mahato, Shiraz Minwalla, Takeshi Morita and Sunil Mukhi for
discussions and comments.

\appendix

\section{\label{step-gap}Some steps for the gap equation}

Let us combine the flavour and spinor indices to write a
four-component fermion
\[
\Psi_i = \left(\begin{array}{c} \psi_{1i} \\ \psi_{2i}
\end{array} \right)
\]
In this notation, \eq{auxil} reads:   
\be
{\cal L} = \int d^3\vec  x\ dt 
\Psi^\dagger_i \tilde D \Psi_i
\label{auxil-gamma}
\ee
where
\be
\tilde D \equiv i\del_t {\bf 1} \otimes {\bf 1}
+ i \del^2 \del_i \s_3 \otimes \s_i 
+  \left( \p^* \s^+ +  \p\ \s^-\right)  \otimes {\bf 1} 
\label{tilde-d}
\ee
We find that subsequent calculations get considerably
simplified if we write the operator $\tilde D$ in terms of Dirac's
gamma matrices $\g^0, \g^i$
\bea
\tilde D &&= \g^0 D
\nn
D &&=  i \g^0 \del_t + i  \g^i \del_i (i\del)^2 
+ \left( \p_R   - i \p_I \g^5 \right)
\label{app-dirac}
\eea
Here $\p = \p_R + i \p_I$.  In our convention
\[
\gamma^0 = \s_1 \otimes {\bf 1}, \g^i = i\s_2 \otimes \s_i, \g^5 =
i\g^0 \g^1 \g^2 \g^3 = \s_3 \otimes {\bf 1}
\]
We emphasize that although we find it expedient to use the
gamma matrices, the operator $D$ above is {\em not} the Dirac
operator. For instance, the coefficient of $\g^i$ has three
powers of momenta, as appropriate for a $z=3$ theory. 

It is obvious that integrating the fermions out from \eq{auxil-gamma}
leads to the effective action \eq{s-eff}.  Let us consider the
equation of motion $\delta S_{\rm eff}/\delta \p_R =0$. This gives

\be
\frac2{g^2}  \p_R = -i N {\rm Tr} \left( \tilde D^{-1} \g^0 \right) 
= -i N {\rm Tr} \left( D^{-1} \right) 
\label{tmp}
\ee
The operator $i D^{-1}$ is simply the propagator. In the momentum
basis it is given by
\[
D^{-1} = 
\frac{k_0 \g^0 + k^2 k_i \g^i -  \left(\p_R + i\p_I
\g^5\right)}{k_0^2 - k^6 -  \p^* \p + i\ep}
\]
Eqn. \eq{gap} now simply follows from \eq{tmp}.

In $d=2n+1$ spatial dimensions, the propagator is $iD^{-1}$, with
\bea
D &&=  i \g^0 \del_t + i \del^{2n} \del_i \g^i 
+ \left( \p_R   - i \p_I \g^{d+2} \right)
\nn
\gamma^0 &&= \s_1 \otimes {\bf 1}, \g^i = i\s_2 \otimes \G_i, \g^5 =
i^n \g^0 \g^1... \g^d = \s_3 \otimes {\bf 1}
\label{gamma-d}
\eea

\section{\label{step-2pt}One loop boson propagator}

In this section we will show the masslessness of the
pion by an explicit one-loop computation.

We will find it convenient, for the purpose
of this calculation, to expand the scalar field $\p$ as 
\[
\p = m^3 + g\e, \, \e = \frac{\ts + i\tp}{\sqrt 2}
\]
To this order, the $\ts$ and $\tp$ fields are simply
the $\s$ and $\pi$ fields of Section \ref{sec:fluct},
up to constant factors. 

Using the form of the action as given by \eq{auxil-gamma} and 
\eq{app-dirac}, we get
\bea
S&&= \int d^4x \left[
\bar\Psi_i \left( i \g^0\del_t + i(\vec \g.\vec \del)(i \vec \del)^2
+ m^3 \right) \Psi_i \right.
\nn
&& ~~~~~~~~~\left. + \frac{g}{\sqrt 2} \bar\Psi_i \Psi_i \ts +  
\frac{g}{\sqrt 2} \bar\Psi_i \g^5 \Psi_i \tp - \frac12
\left((\frac{m^3\sqrt 2}{g} + \ts)^2 + \tp^2\right)
\right]
\label{fluct-action}
\eea

\subsection{Summary of results}

The tree-level propagator for $\ts$ and $\tp$ fields are
non-dynamical. However, the propagators develop non-trivial correction
through fermion loops. We present the summary of results here and
defer details of the computation to the next subsection.  To leading
order in $1/N$, we find the following results for the propagators
$G_\ts(p)$ and $G_\tp(p)$ for the $\ts$ and $\tp$ fields, respectively:
\beas
G_\ts(p) &&=  \frac{-i}{1+ i\G^{(2)}_\ts(p)},\;\G^{(2)}_\ts(p)= 
i\left(1- \frac\l {6\pi^2}\right) + o(p^2)
\nn
G_\tp(p) &&=  \frac{-i}{1+ i\G^{(2)}_\tp(p)},\; \G^{(2)}_\tp(p)= 
i + o(p^2)
\eeas 

In the small $p$ limit, 
\be
G_\ts(p)= \frac{1}{\l/(6\pi^2) + o(p^2)},\; G_\tp(p)= \frac{-i}{o(p^2)}
\label{sig-prop}
\ee
Therefore, the pion propagator has a massless pole, whereas the $\ts$
field is massive.

\subsection{Details}

The Feynman rules that follow from \eq{fluct-action} are given by:

\gap2

\ni Fermion propagator: \kern5pt\lower9pt\hbox{
\includegraphics[scale=.4]{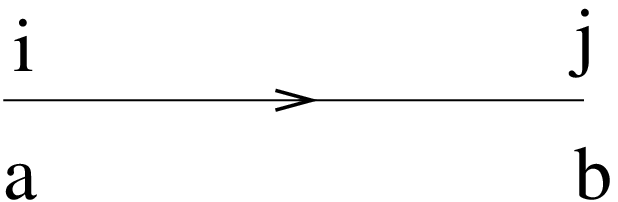}}
\kern22pt
$= \frac{i\delta_{ab}\delta_{ij}}{\g^0 p_0 + \vec \g. \vec p \ p^2 + m^3}= 
\Delta_F(p)$ 

\gap3

\ni Yukawa couplings: \kern15pt\lower20pt\hbox{
\includegraphics[scale=.4]{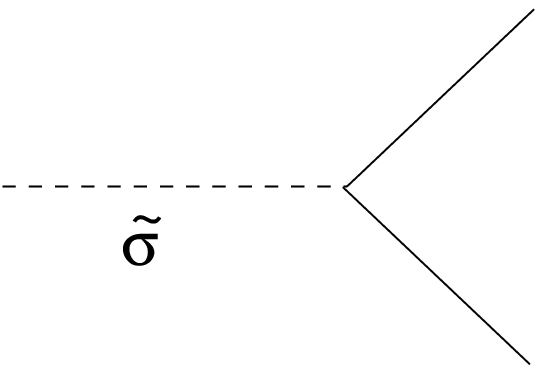}}\kern35pt $= \frac{ig}{\sqrt 2}$

\gap3

\ni \kern115pt\lower9pt\hbox{
\includegraphics[scale=.4]{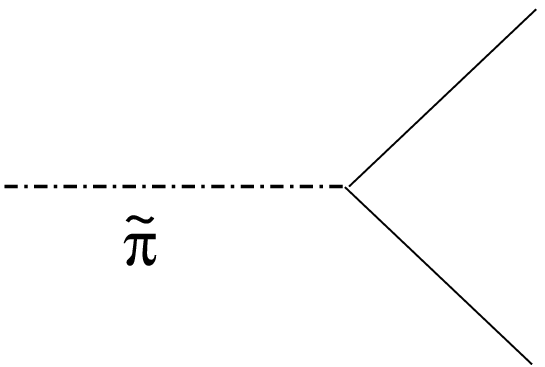}}\kern35pt 
$= \frac{-g\g^5}{\sqrt 2}$

\gap2

\ni The propagators for $\ts$ and $\tp$ are simply given by $-i$.

\gap3

We will first compute the one-loop two-point function of $\ts$. To
order $g^2$, it is represented by the following Feynman diagram

\includegraphics[scale=0.6]{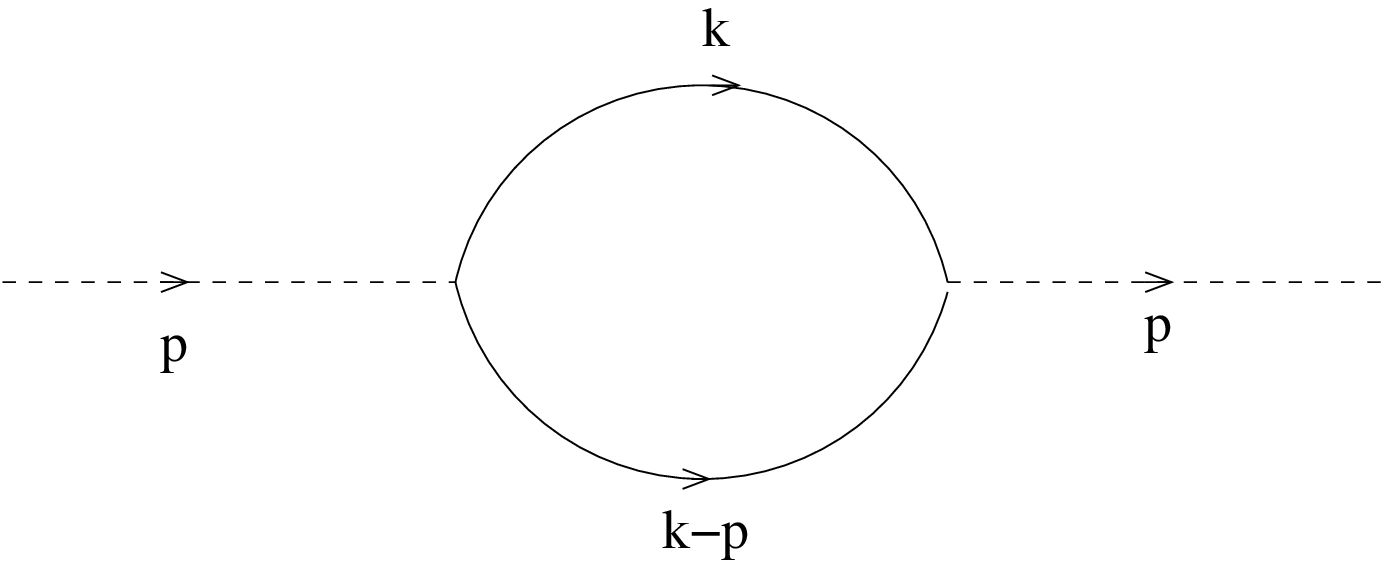}

\ni which evaluates to
\bea
\G^{(2)}_\ts(p)\mm= (-1)\ {\rm Tr} \int \frac{d^4k}{(2\pi)^4}
\frac{ig}{\sqrt 2} \Delta_F(k)\ \frac{ig}{\sqrt 2} \Delta_F(k-p) 
\nn
\mm= -2\l  \int \frac{d^4k}{(2\pi)^4} 
\frac{k_0(k_0-p_0) - \vec k.(\vec k - \vec p) (\vec k)^2( \vec k - \vec p)^2
+ m^6}{(k_0^2 - \vec k^6 - m^6)((k_0 - p_0)^2 - ( \vec k - \vec p)^6 - m^6)}
\label{sig-2pt}
\eea
The full propagator for $\ts$ at momentum $p$ is obtained by summing over
an infinite series of such diagrams, and we obtain
\[
G_\ts(p)= -i + (-i)\G^{(2)}_\ts(p)(-i)+ ...= \frac{-i}{1+ i\G^{(2)}_\ts(p)}
\]
Note that by changing $k_0 \to - k_0$ and $\vec k \to -\vec k$ we can
prove that $\G^{(2)}_\ts(-p)= \G^{(2)}_\ts(-p)$. Note also that 
\beas
\G^{(2)}_\ts(0) &&\kern-20pt
= -2\l \left(-\frac{i}{2\l} + 2m^6 \int\frac{d^4k}{(2\pi)^4} 
\frac{1}{(k_0^2 - k^6 - m^6)^2}
 \right)
\nn
&&
= i - 4\l\ m^6\frac{\del}{\del m^6}\left(\frac{-i}{2\l}\right)
\G^{(2)}_\ts(0) = i\left(1- \frac\l {6\pi^2}\right)
\label{1PI-sigma}
\eeas
Hence, at $p\to 0$, 
\[
G_s(0)= \frac{-i}{\l/6\pi^2}
\]
which shows that $\ts$ is a massive particle.

\gap1

The one-loop two-point function for $\tp$ is represented by a
Feynman diagram similar to the above, and is given by
\bea
\G^{(2)}_\tp(p) &=& (-1)\ {\rm Tr} \int \frac{d^4k}{(2\pi)^4}
\frac{-g \g^5}{\sqrt 2} \Delta_F(k)\ \frac{-g \g^5}{\sqrt 2} \Delta_F(k-p) 
\nn
&=& -2\l  \int \frac{d^4k}{(2\pi)^4} 
\frac{k_0(k_0-p_0) - \vec k.(\vec k - \vec p) (\vec k)^2( \vec k - \vec p)^2
- m^2}{(k_0 - \vec k^6 - m^6)((k_0 - p_0)^2 - ( \vec k - \vec p)^6 - m^6)}
\nn
\label{pi-2pt}
\eea

As for $\ts$, the full propagator for $\tp$ at momentum $p$ is given by
the sum
\[
G_\tp(p)= -i + (-i)\G^{(2)}_\tp(p)(-i)+ ...= \frac{-i}{1+ i
\G^{(2)}_\tp(p)}
\]

Note, like before, that $\G^{(2)}_\tp(-p)= \G^{(2)}_\tp(-p)$. Also
$\G^{(2)}_\tp(0) = -2\l \frac{-i}{2\l}=i$. Thus, $\G^{(2)}_\tp(p)= i +
o(p^2)$.

Hence, as $p \to 0$, 
\[
G_\tp(p)= \frac{-i}{o(p^2)}
\] 
Thus, the $\tp$ propagator has a pole at $p^2=0$. Hence the
pion is massless.

\subsection{\label{sec:fermi-2pt}Fermion two-point function}

In this subsection we present an expression for the fermion 2-point
function $\G^{(2)}_F(p)$. To $o(g^2)$, it is given by the following
diagram (the blobs represent the full propagators $G_\ts(p)$ and
$G_\tp(p)$, respectively)

\gap3

%\begin{figure}
\centerline{\includegraphics[scale=.6]{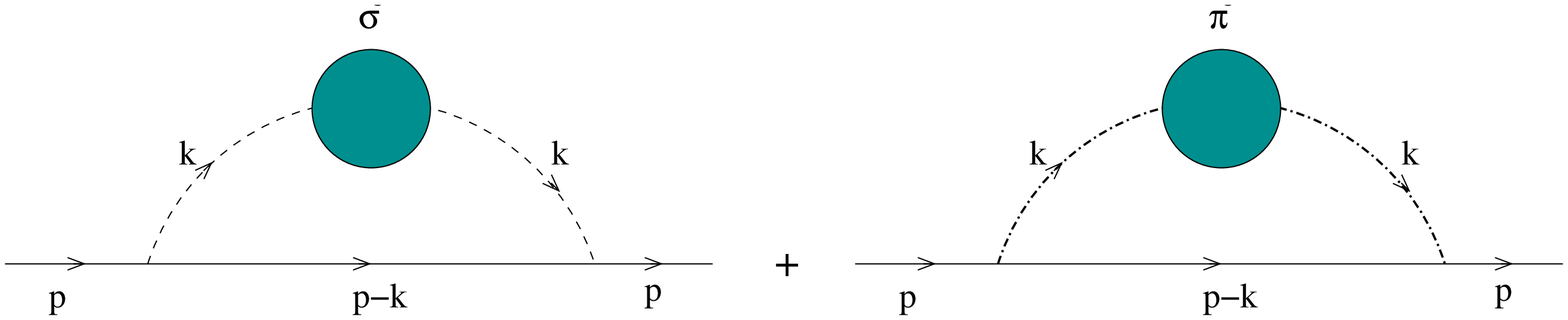}}
%\caption{The blob represents the full propagator $G_\ts(p)$.}
%\end{figure}

\ni The diagram evaluates to
\bea
\kern-15pt\G^{(2)}_F(p) &&\kern-18pt= \int \frac{d^4k}{(2\pi)^4}
\left[ \frac{ig}{\sqrt 2}\ i G_\ts(k) \frac{ig}{\sqrt 2}\ 
\Delta_F(p-k) + \frac{-g}{\sqrt 2}\ i G_\ts(k) \frac{-g}{\sqrt 2}
\ \g^5 \Delta_F(p-k)\ \g^5  \right]
\nn
&&\kern-18pt= \frac{-\l}{2N}
\int \frac{d^4k}{(2\pi)^4} 
\left[ \frac{G_\ts(k+p)}{\g^0k_0 + \vec \g. \vec k\ (\vec k)^2 -m^3}
+  \frac{G_\tp(k+p)}{\g^0k_0 + \vec \g. \vec k\ (\vec k)^2 + m^3}
\right]
\label{generate-g1}
\eea  
The expression, at least formally, contains terms involving
$\vec p. \vec \g$, which renormalize the
relevant coupling $g_1$ in \eq{relevant}. We postpone a 
detailed analysis of this diagram to future work.

\end{document}